\newcounter{myctr}

\documentclass{ws-acs}

\usepackage[compress]{cite}
\usepackage{xcolor}
\usepackage[verbose]{hyperref}
\hypersetup{colorlinks=false,allbordercolors=blue,pdfborderstyle={/S/U/W 1}}
\usepackage[colorinlistoftodos]{todonotes}

\usepackage{algorithm}
\usepackage{algpseudocode}

\usepackage{makecell}
\usepackage{multirow}

\begin{document}

\makeatletter
\def\@biblabel#1{[#1]}
\makeatother

\markboth{L. Anne et al.}{RECCS Synthetic Network Generator}

%
\catchline{}{}{}{}{}
%

\title{RECCS: Realistic Cluster Connectivity Simulator  for Synthetic Network Generation}

\author{Lahari Anne}

\address{Siebel School of Computing and Data Science, University of Illinois Urbana-Champaign, \footnote{Urbana, Illinois 61801, United States of America}\\
\email{lanne2@illinois.edu}}

\author{The-Anh Vu-Le}

\address{Siebel School of Computing and Data Science, University of Illinois Urbana-Champaign, \footnote{Urbana, Illinois 61801, United States of America}\\
\email{vltanh@illinois.edu}}

\author{Minhyuk Park}

\address{Siebel School of Computing and Data Science, University of Illinois Urbana-Champaign, \footnote{Urbana, Illinois 61801, United States of America}\\
\email{minhyuk2@illinois.edu}}

\author{Tandy Warnow}

\address{Siebel School of Computing and Data Science, University of Illinois Urbana-Champaign, \footnote{Urbana, Illinois 61801, United States of America}\\
\email{warnow@illinois.edu}}

\author{George Chacko}

\address{Siebel School of Computing and Data Science, University of Illinois Urbana-Champaign, \footnote{Urbana, Illinois 61801, United States of America}\\
\email{chackoge@illinois.edu}}
\maketitle

\begin{history}
\received{(Day Month Year)}
\revised{(Day Month Year)}
\accepted{(Day Month Year)}
\published{(Day Month Year)}
\end{history}

\begin{abstract}
The limited availability of useful ground-truth communities in real-world networks presents a challenge to evaluating and selecting a ``best'' community detection method for a given network or family of networks. The use of synthetic networks with planted ground-truths is one way to address this challenge. While several synthetic network generators can be used for this purpose, Stochastic Block Models (SBMs), when provided input parameters from real-world networks and clusterings, are well suited to producing networks that retain the properties of the  network they are intended to model. We report, however, that SBMs can produce disconnected ground truth clusters; even under conditions where the input clusters are connected. In this study, we describe the REalistic Cluster Connectivity Simulator ({\sc RECCS}), which, while retaining approximately the same quality for other network and cluster parameters, creates an SBM synthetic network and then modifies it to ensure an improved fit to cluster connectivity. We report results using parameters obtained from clustered real-world networks ranging up to 13.9 million nodes in size, and demonstrate an improvement over the unmodified use of SBMs for network generation.
\end{abstract}

\section{Introduction}

 Community detection methods uncover the meso-scale structure of networks, where communities are groups of nodes that display distinct structural and connectivity patterns. These groups or clusters are typically characterized by high internal density, separation from the rest of the network,  and strong internal edge connectivity \cite{Fortunato2022,ElMoussaoui2019,Traag2019,Park2024}. 
 
A  large number of community detection methods are available \cite{Javed2018,Coscia2011}. However, selecting a community detection method for a given application can be challenging since real-world networks are rarely annotated with relevant ``ground truth" communities. Instead, researchers rely on synthetically generated networks with planted ground-truths \cite{Orman2013,Peel2017,Zhang2020} that enable comparisons between clustering methods. For this purpose, several synthetic network generators, such as Stochastic Block Models (SBM) \cite{peixoto_graph-tool_2014,Peixoto-chapter-2019}, LFR \cite{Lancichinetti2009}, ABCD and ABCD+o \cite{abcd,abcdo}, and nPSO \cite{Muscoloni2018}, have been developed. 
 
Synthetic networks should exhibit the structural and clustering properties of the real-world networks they are intended to model \cite{vaca2022systematic,Muscoloni2018}. In this respect, Viger and Latapy \cite{viger2016efficient} achieved efficiency in modeling degree sequences and generating connected graphs but did not address community structure. Deep learning approaches such as GraphRNN \cite{graphRNN} face challenges in scaling to large networks. 
Stochastic Block Models (SBMs) implemented in graph-tool \cite{peixoto_graph-tool_2014}---referred to as SBMs hereinafter---
aim to enable the reproduction of intra- and inter-cluster relationships and  have been shown to be  highly scalable  \cite{Peixoto-chapter-2019}. 
Furthermore, SBMs have been shown to replicate key network characteristics, such as degree distribution, clustering coefficients, and network diameter \cite{vaca2022systematic}. 
Nevertheless, limited work has been conducted on assessing how well the ground truth clusterings generated by SBMs align with those in real-world networks, including whether SBMs accurately reflect portions of the network being unclustered (equivalently, nodes that are in singleton clusters, and can be referred to as outlier nodes).  

In a prior study \cite{anne2024}, we  reported that SBMs often produce disconnected ground truth clusters, even when the input parameters are obtained from networks clustered using methods that are guaranteed to produce connected clusters. 
To address this limitation, we introduced the ``REalistic Cluster Connectivity Simulator" ({\sc RECCS}), a two-step framework that remediates the edge connectivity of clusters generated by SBMs and   generally preserves other network properties. 

Given a clustered network, {\sc RECCS} comprises two steps.  In the first step, it produces an SBM that models the clustered portion of the network, aiming to reproduce the edge connectivity of the input network.  In the second step, outlier nodes are added to the SBM.  

In \cite{anne2024}, we examined two ways of accomplishing the first step and three ways of performing the second step.
We found that both ways of varying Step 1, which models the clustered subnetwork, were  improvements over a naive use of the SBM software; in addition, each had strengths relative to the other, making them both of further interest.
We also compared three ways of adding the unclustered (outlier) nodes into the SBM network produced during the first step, and 
found that only one provided excellent fit to the real-world network statistics, 

In this extended version of  \cite{anne2024}, we provide an extended study on {\sc RECCS}, as follows.
Leveraging results from \cite{anne2024}, we focus on just one way of handling outliers, so that we propose and study two variants of {\sc RECCS} that differ only in how they model the clustered subnetwork.
We  performed additional experiments to evaluate these two versions of {\sc RECCS}, and include  results on nine new clustered networks for input parameters.
For these new test networks,  the clustering method used to provide the input parameters includes the  
Connectivity Modifier (CM) \cite{Park2024}, which post-processes clusters to enforce well-connectedness, on the input clustering used to generate SBMs. 
We evaluate clustering methods for accuracy with respect to recovery of ground truth clusters using  synthetic networks generated by {\sc RECCS}.
We  conduct replication experiments, testing {\sc RECCS} across multiple runs with the same and different SBM base networks. 
Finally, we provide runtime metrics for {\sc RECCS} under various  conditions.

\section{Materials and Methods}

\subsection{\sc{RECCS} pipeline}\label{sec:reccs-workflow}

We use {\sc RECCS} within a pipeline that begins with a real-world network $G$ and a clustering $\mathcal{C}$ of  $G$. 
Any node in the graph that is not in a cluster of size at least two is considered an ``outlier" node, and the other nodes are considered ``clustered" nodes. The subnetwork induced by the clustered nodes, and so containing only those edges that go between clustered nodes, is referred to as $G_c$.

We compute the edge connectivity of each cluster,  defined as follows.
Given a cluster $C$ of nodes in $G$, an
 {\em edge cut} for $C$ is  a subset of edges connecting nodes in $C$ so that when that set of edges is removed, $C$ splits into two or more parts.
 The number of edges in  the smallest edge cut is the edge connectivity of the cluster $C$.
 Thus, if $C$ is internally disconnected, its edge connectivity is $0$, and otherwise the edge connectivity is at least $1$.
 We refer to the edge connectivity of cluster $C$ by $k(C)$.
 Given the context, we often refer to this as the ``connectivity" of $C$.

Given the clustering $\mathcal{C}$ of the network $G$, we compute $G_c$, and for this subnetwork we compute  the   degree sequence, the assignment of nodes to non-singleton clusters, the edge count matrix that defines the number of edges within each cluster and between every pair of clusters, and the edge connectivity of each cluster.
We refer to this set of parameters  by $Param(G_c,\mathcal{C})$.

In our experiments, we have noticed that SBMs often produce extra edges in the form of self-loops and parallel edges. Since our objective is to return a simple graph, we always follow the construction of an SBM by a step where we remove the self-loops and excess edges. This gives us the opportunity to add back edges that improve the fit to the cluster edge connectivity and other parameters, as desired. 

\begin{figure}[t]
\centerline{\includegraphics[width=1\textwidth, height=0.4\textheight]
{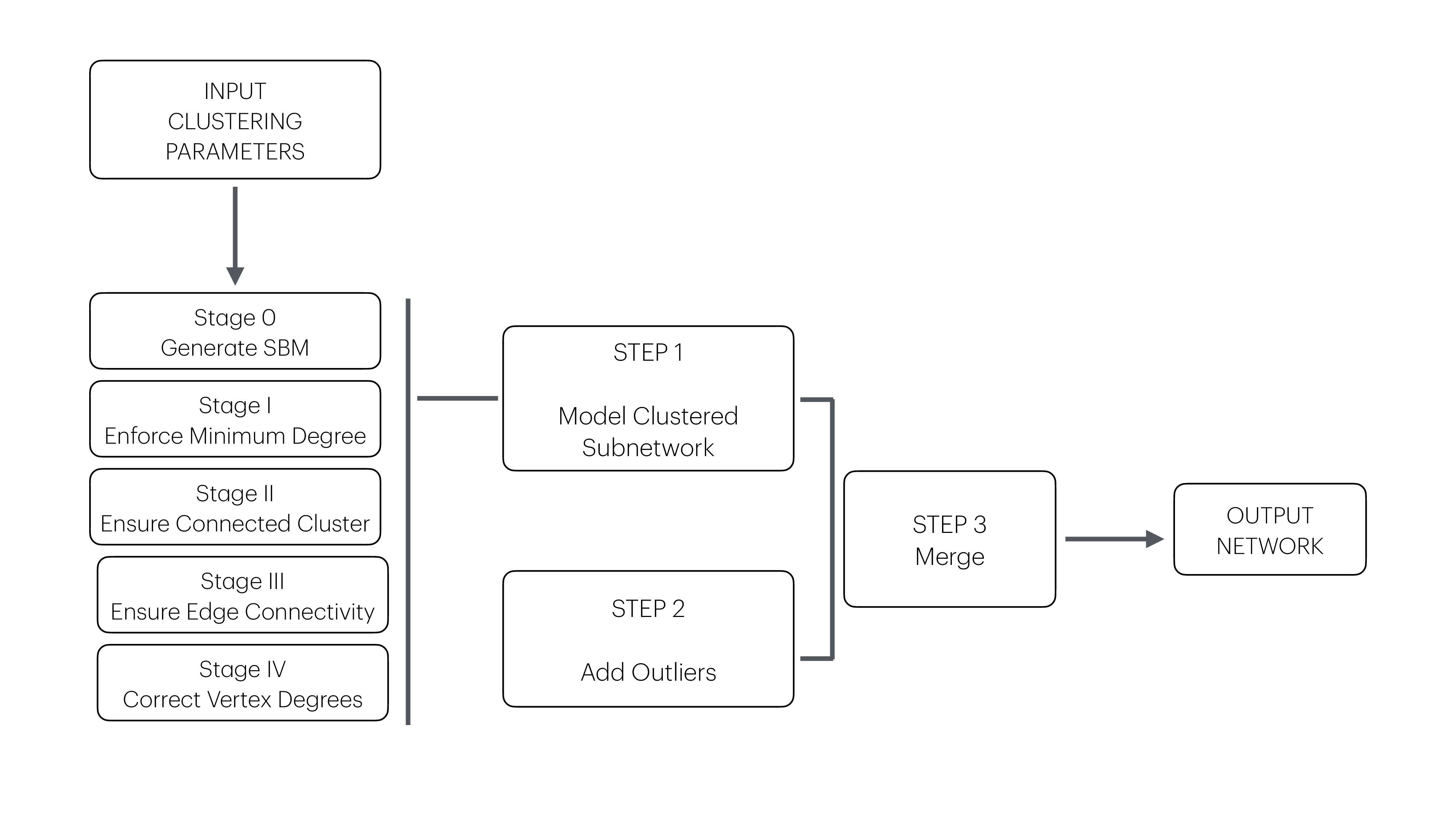}}
\caption{\textbf{{\sc RECCS} Overall Workflow.} The workflow modifies an initial network by adding edges to improve its fit to the input parameters. The input to {\sc RECCS} consists of parameters derived from a real-world network $N$ and its estimated clustering. 
Step 1 generates a synthetic network representing the clustered subnetwork of $N$, while Step 2 generates a network modeling the ``outlier" nodes. 
The two networks are then merged in Step 3.
For more details, see Section \ref{sec:reccs-workflow}.}
\label{fig_ce_flow}
\end{figure}

The algorithmic structure of {\sc RECCS} (Figure \ref{fig_ce_flow}) has three  steps. Step 1 produces a synthetic network that models the clustered subnetwork so that all edges are between clustered nodes, and Step 2 produces a synthetic network that addresses the outliers. 
In Step 3, the two networks from Steps 1 and 2 are then superimposed to produce the final returned network.

\subsubsection{Step 1 of {\sc RECCS}: Modeling the Clustered Subnetwork}

Step 1 of {\sc RECCS} has five  stages, where  the first four stages are concerned with edge connectivity  and the final stage is concerned with improving the fit to the degree sequence.
The high-level description of Step 1 is as follows. 

\begin{itemize}
\item \textbf{Stage 0: Construct SBM.}
We construct an SBM under the degree-corrected model for the parameter set 
$Param(G_c,\mathcal{C})$ using graph-tool.  We then  remove excess edges. 
We refer to this network as $N_c$.
Note that this network models the clustered subnetwork $G_c$ of $G$.
\item \textbf{Stage 1: Enforce minimum degree.} Here we add edges to ensure that every node has  at least $k(C)$ neighbors in the cluster. 
Specifically, for a cluster with a required minimum cut size of $k(C)$, each node within the cluster should have at least $k(C)$ edges to nodes within the same cluster. Therefore, if a node $v$ has $d < k(C)$ neighbors in the cluster, we add $k(C)-d$ edges between $v$ and other nodes in the cluster to which it is not adjacent.

\item \textbf{Stage 2: Ensure connected clusters.} If a cluster $C$ is disconnected, we add $k(C)$ edges at random between its largest component and each of the other components.

\item \textbf{Stage 3: Ensure edge connectivity at least $k(C)$. } 
This stage is an iterative method that ensures that the cluster $C$ has edge connectivity at least $k(C)$. Specifically, we use VieCut \cite{henzinger2018practical} to calculate the size of a minimum edge cut; if this size is at least $k(C)$, then we stop (the cluster is sufficiently well connected).
Otherwise, we identify the two parts of the cluster that are connected by fewer than $k(C)$ edges, and we add edges between the two parts to achieve  $k(C)$.  We then repeat the process until the minimum cut size of the cluster is greater than $k(C)$.

\item \textbf{Stage 4: Improve vertex degree fit.}
During this stage, edges are added to improve the fit to the target degree sequence.

 \end{itemize} 

During the development of {\sc RECCS}, we explored two different techniques for Stage 4; the results of 
these explorations are reported in Section \ref{sec:exp2}, and show that both techniques were reasonable, each providing an improvement over the initial SBM but neither dominating the other.

\subsubsection{Step 2 of {\sc RECCS}: Modeling Outliers}\label{outlierstrategies} 

 Recall that we begin with a given real-world  network $G$ and its clustering $\mathcal{C}$,  and that Step 1 produces a synthetic network to model the clustered subnetwork.
 In Step 2 we wish to add the outlier nodes into the synthetic network, where the outlier nodes are those that were not in any non-singleton cluster.

 We achieve this by creating a synthetic for the subnetwork $G^*$ that includes only edges involving at least one outlier node, referred to as ``outlier edges." These edges either connect two outlier nodes or an outlier node with a clustered node. 

 Importantly, $G^*$ includes both clustered and outlier nodes and retains the same cluster structure as $\mathcal{C}$. However, there are no edges within any non-singleton cluster or between non-singleton clusters in $G^*$.

We create a synthetic network $N^*$ using the parameters for $G^*$ as follows. Each outlier node is treated as a singleton cluster. Thus, if there are $k$ outliers and $L$ non-singleton clusters, the SBM input will include $L + k$ clusters in total. The edge count matrix specifies the number of edges between any two clusters (including the singleton clusters).  Note that the edge count matrix entry for two singleton clusters is either $1$ or $0$, depending on whether the two nodes are adjacent.    

Note that this method exactly reproduces the edges between outliers (as they belong to singleton clusters) and matches the specified number of edges between singleton clusters (outliers) and non-singleton clusters. However, edges between outlier nodes and clustered nodes are randomly assigned. We then remove all excess edges (self-loops or parallel edges), thus producing a simple graph that models the outliers. 

\subsubsection{Step 3 of {\sc RECCS}: Merging }\label{merging}

Finally, the synthetic networks generated in Steps 1 and 2 are combined into a single network. This merging process is straightforward, as the node labelings in both synthetic networks are consistent with those in $G$.

\subsection{Experimental Study}
We generated synthetic networks using variants of RECCS and SBM, each based on
empirical statistics  from clustered real-world networks.
Each clustered real-world network was used either in a development phase or in a testing phase, and is  labelled as such.

We computed the similarity of RECCS networks to the clustered real-world networks using a number of different statistics. We also report performance data of RECCS. Finally, we studied the accuracy of different clustering methods on RECCS synthetic networks using the NMI and ARI accuracy measures. 

\subsection{Datasets}
\label{sec:datasets}

The development dataset has 110 clustered networks and the testing dataset consists of 9 networks clustered using a variety of methods. These datasets encompass a wide variety of real-world network types.

\subsubsection{Clustering}

Each real-world network was clustered  using several methods, including the Leiden algorithm \cite{Traag2019} optimizing either the Constant Potts Model (CPM) with a range of resolution parameters or optimizing Modularity, the Iterative k-core method  (IKC) \cite{wedell2022center}, or Infomap \cite{infomap}. 
Each of these clustering methods is either guaranteed to produce connected clusters (e.g., Leiden) or post-processed to return the components if they are not so guaranteed (e.g., IKC and.Infomap).
In some cases, these clustering methods were followed by a post-processing technique to ensure that the clusters were well-connected.
To achieve this, we used the Connectivity Modifier (CM) \cite{Park2024}, which enforces an edge-connectivity of greater than $\log_{10}(n_C)$, where $n_C$ represents the number of nodes in the cluster. 
Clustering was performed both with and without the CM post-processing step. 

During the development phase, we  filtered out (i.e., removed) very small clusters (those having at most 10 nodes) when using CM. During the evaluation phase, we did not filter out the small clusters when using CM; this change allowed us to avoid overlooking any effects restricted to very small clusters.

\subsubsection{Primary dataset: 110 real-world networks}

The primary dataset consists of \textit{110 undirected real-world networks}, of which  104 networks were sourced from the Netzschleuder catalog \cite{Netzschleuder}. 
These 104 networks vary in size, ranging from 900 to 1402673 nodes, and 2400 to 17,233,144 edges.

Six other networks were included, and are listed below; these were used as benchmarks by us \cite{Park2024}, with  five  available in the SNAP repository \cite{leskovec2016snap} and the sixth generated by us \cite{wedell2022center}. 

\begin{itemlist}
\item Cit-Hepph 
\item Cit-Patents 
\item Orkut 
\item Wiki-Topcats
\item Wiki-Talk
\item Curated Exosome Network (CEN)
\end{itemlist}

These six networks vary in size, with node counts ranging from 34,546 to 13,989,436 and edge counts ranging from 420,877 to 117,185,083.

\subsubsection{Development dataset}
For development, all 110 networks in the primary dataset were clustered using Leiden optimizing CPM with a resolution parameter $r=0.001$ \cite{Traag2019}, followed by  post-processing (see above) to ensure that all clusters are well-connected, and also all clusters with at most 10 nodes are deleted.
These clustered networks were then used during the algorithm design phase.

\subsubsection{Testing Set}

We have two collections of clustered networks, Testing 1 and Testing 2, that we used in our evaluation of {\sc RECCS}v1 and {\sc RECCS}v2, in comparison to SBM.

The real-world networks for Testing 1 are the same six benchmark networks from the SNAP repository from the primary dataset, but clustered differently than for the Development dataset.
These networks were clustered using Leiden optimizing CPM ($r=0.01$; \emph{a different resolution} from the $r=0.001$  
used during development), Leiden optimizing Modularity, and Iterative k-Core (IKC) \cite{wedell2022center} with $k=10$, followed by post-processing (see above).

The real-world networks for Testing 2  are taken  from SNAP but are not in the primary dataset: 
\begin{itemize}
\item gemsec-Facebook (artist edges) 
\item com-Amazon
\item com-LiveJournal (com-lj edges) 
\item com-Youtube
\item gemsec-Deezer (HR edges) 
\item twitch-gamers (large twitch edges)
\item musae-github
\item soc-Pokec
\item ego-Twitter
\end{itemize}

These networks range in number of nodes ranging from 37,700 to nearly 4 million and number of edges ranging from 289,003 to over 34 million. Each network in this dataset was clustered using multiple methods, including Leiden optimizing CPM (with a resolution parameter $r=0.01$), Leiden optimizing Modularity, Iterative k-core (IKC) with $k=10$, and Infomap \cite{infomap}. 
These clusterings were performed both with and without  the post-processing step described above.

\subsection{Evaluation Criteria}\label{eval_criteria}

We evaluate the fit between the synthetic network with its associated ground truth clustering and the input clustered real-world network using a combination of scalar and sequence statistics computed on these networks given in Table \ref{tab:network_properties}. Some statistics, such as degree sequence, global and local clustering coefficients, and diameter describe network properties, while others, such as minimum edge cut size and mixing parameter, concern cluster properties.

\begin{table}[h]
\tbl{Key Network Properties and Their Distance Metrics.\label{tab:network_properties}}
{\begin{tabular}{@{}ccc@{}} \toprule
Statistic & Type of Statistic & Distance Metric \\ \colrule
Minimum Edge Cut Size & Sequence & RMSE \\
Diameter & Scalar & Relative Difference \\
Mixing Parameter & Scalar & Simple Difference \\
Degree & Sequence & RMSE \\
Global Clustering Coefficient & Scalar & Simple Difference \\
Mean Local Clustering Coefficient & Scalar & Simple Difference \\
Edges between Outliers \& Clustered Nodes & Scalar & Relative Difference \\
Outlier Node Degree & Sequence & RMSE \\ \botrule
\end{tabular}}
\begin{tabnote}
This table outlines the network properties and the metrics used to evaluate the similarity between real-world and synthetic networks.
\end{tabnote}
\end{table}

We denote a statistic for the real-world network by $s$ and its corresponding statistic for the synthetic network by $s'$. For statistics that are scalars (i.e., real-valued) and bounded between $0$ and $1$, we use the simple difference (Eq. (\ref{eq:diff})). For scalar statistics that are not bounded by $0,1$, we use the relative distance (Eq. (\ref{eq:reldiff})).

\begin{equation}
    \text{Simple Difference} = s - s' \label{eq:diff}
\end{equation}

\begin{equation}
    \text{Relative Difference} = \frac{s - s'}{s}
    \label{eq:reldiff}
\end{equation}

For those statistics that are sequences (e.g., the degree sequence, minimum edge cut size for the clusters), we let the $i^{th}$ entries be denoted by $s_i$ and $s'_i$, and we assume that the sequences are of length $n$.
For these comparisons, we use the \emph{Root Mean Square Error (RMSE)}, given by Eq. (\ref{eq:rmse}).

\begin{equation}
    \text{RMSE} = \sqrt{\frac{1}{n} \sum_{i=1}^{n} (s_i - s'_i)^2} 
    \label{eq:rmse}
\end{equation}

Finally, we also evaluate the fit between two networks using a holistic criterion: the normalized edit distance, which is the number of edges in one network and not in the other. This is enabled since there is a bijection between the node sets of the synthetic networks and the real-world networks.  

\section{Results and Discussion}

Results are presented in the following six sections.

\begin{itemize}
    \item Experiment 1: Evaluating SBM networks
    \item Experiment 2: Designing  RECCS  
    \item Experiment 3: Evaluating RECCS  
    \item Experiment 4: Evaluating clustering methods on RECCS networks
    \item Experiment 5: Replication
    \item Experiment 6: Evaluating computational performance for RECCS
\end{itemize}

\subsection{Experiment 1: Evaluating SBM networks}\label{sec:exp1}

\begin{figure}[t]
\centerline{\includegraphics[width=0.9\textwidth]{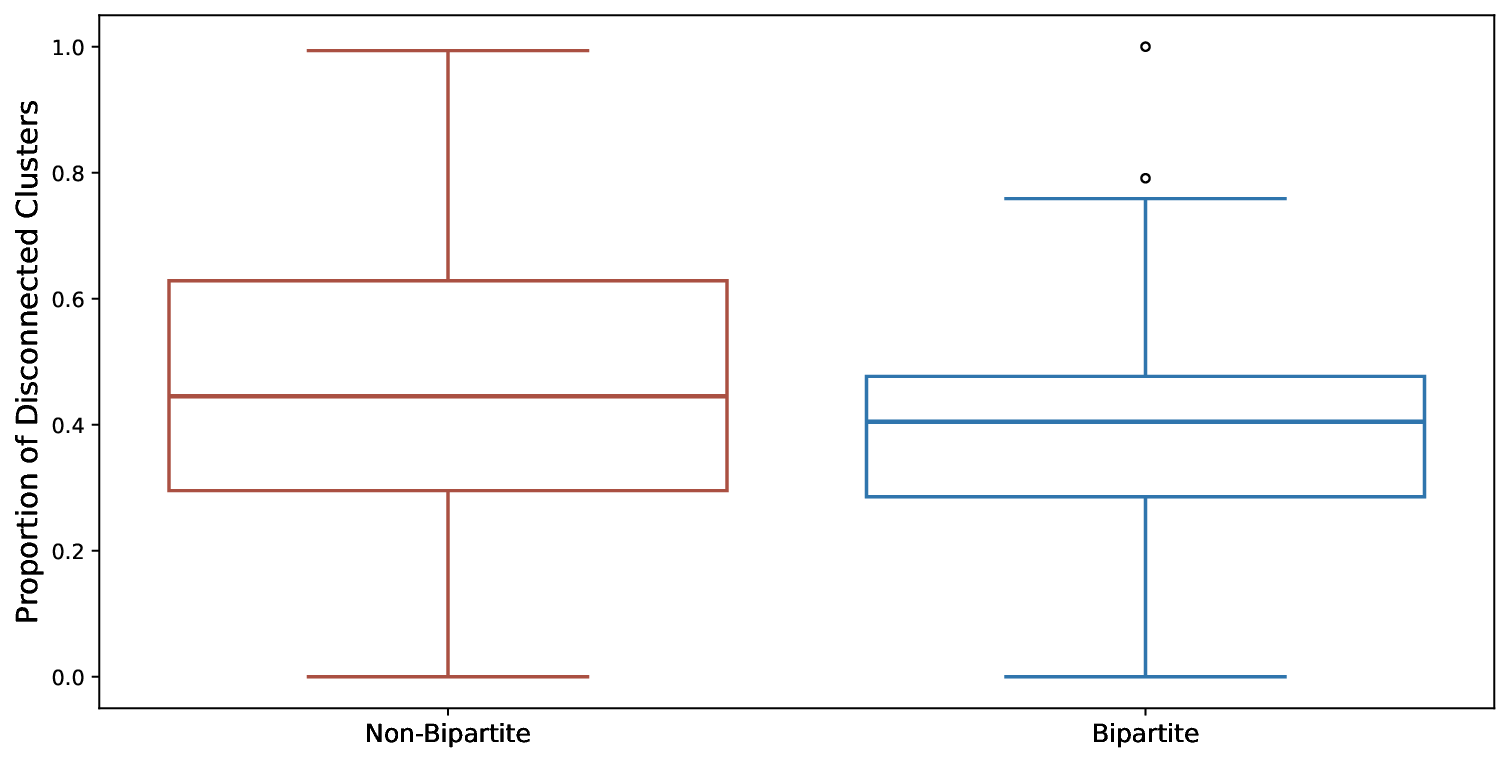}}
\vspace*{8pt}
\caption{\textbf{Proportion of disconnected clusters observed in SBM synthetic networks}. 
There are 30 bipartite networks ranging from 135 to 302,745 nodes and 80 non-bipartite networks ranging from 552 to 337,7185 nodes, both size ranges reported for \emph{clustered subnetworks}. SBMs  were created using the Development dataset, which are based on parameters from these real-world networks clustered with the Leiden algorithm optimizing CPM at $r=0.001$ + CM with small clusters removed.
This figure demonstrates that SBMs synthetic networks have many disconnected clusters (considering only non-singleton clusters) for these conditions, and for both bipartite and non-bipartite networks.}
\label{fig_disconnect_proportions}
\end{figure}

 We evaluated SBMs generated under the degree-corrected model on the Development datasets, which are  clustered real-world networks where all the clusters are guaranteed to be well-connected (Section \ref{sec:datasets}). Strikingly, more than 50\% of these SBM networks contain at least 40\% disconnected clusters, and 
both bipartite and non-bipartite networks show this trend (Figure \ref{fig_disconnect_proportions}).
 Thus, ground truth clusters in SBM synthetic networks generated from clustered real-world networks do not meet minimal expectations of edge-connectivity for clusters.
 This finding is in contrast to the high accuracy of SBMs with respect to network properties such as diameter, degree sequence, and clustering coefficient,  reported in \cite{vaca2022systematic}.  These observations motivate the development of {\sc RECCS} and the remainder of this study.

\subsection{Experiment 2: Designing {\sc RECCS}}\label{sec:exp2}

During the development of {\sc RECCS}, earlier designs omitted Stage 4, where edges are added to improve the fit to the degree sequence.
These initial designs achieved strong results for edge connectivity but performed poorly in matching the degree sequence.  Step 1 was then modified in order to improve the fit to  the degree sequence, leading to two distinct versions of {\sc RECCS}, differing only in the implementation of Stage 4.

 \begin{figure}[t]
\centerline{\includegraphics[width=\textwidth]{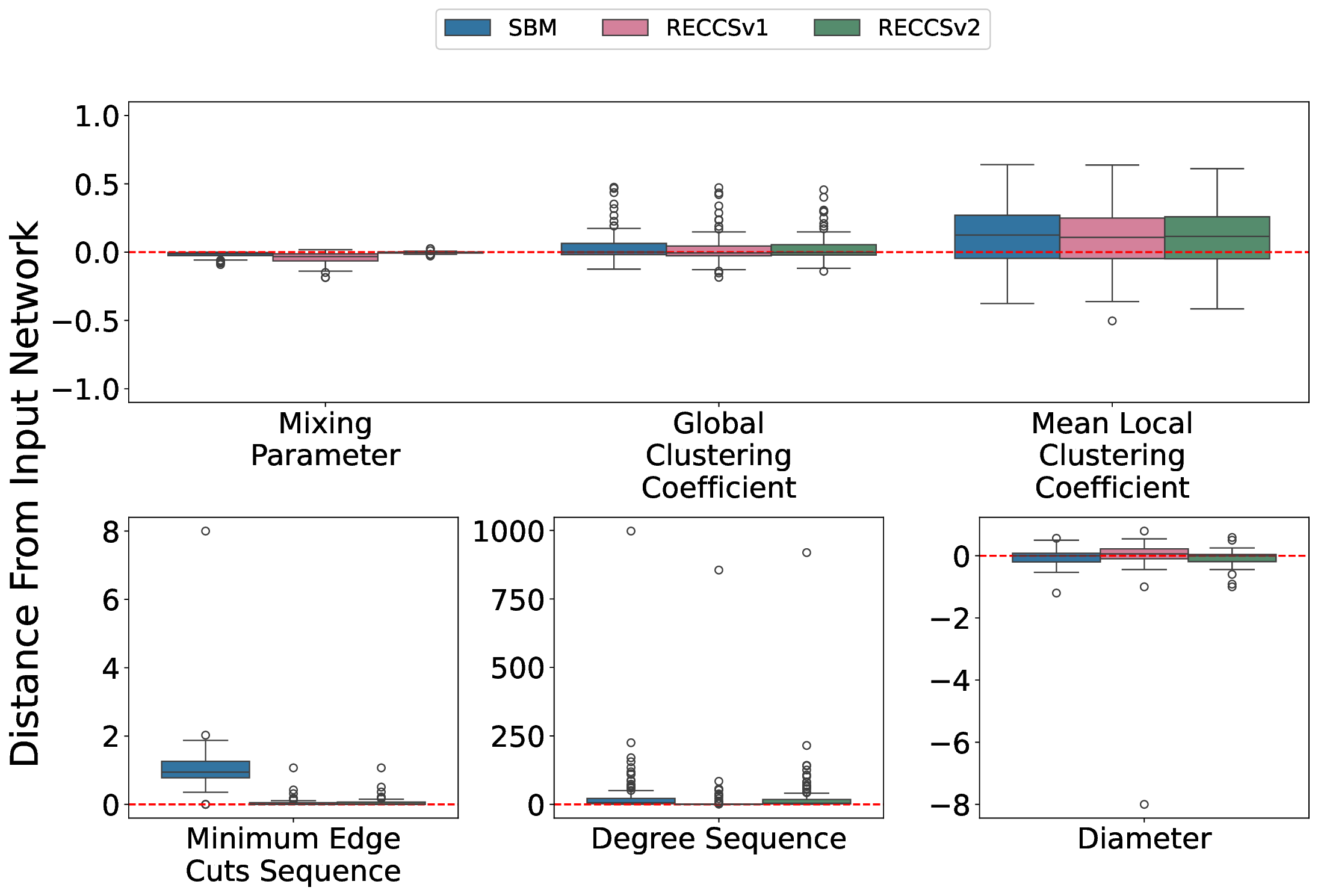}}
\vspace*{8pt}
\caption{\textbf{Experiment 2: Comparing  {\sc RECCS} (v1 and v2) to  SBM networks  with respect to similarity to the clustered subnetworks}. Both RECCS v1 and v2 show improvements over cluster edge-connectivity compared to  SBM networks. The y-axis lists various distance metrics for network properties. 
For clustering coefficients, ratios, and mixing parameters, the simple scalar difference is reported. Relative differences are shown for network diameter, and RMSE is used for the degree sequence and minimum edge cut sequence.  
This experiment was conducted on the development networks. }
\label{fig_ce_boxplot}
\end{figure}

Recall that we are given a target degree  for every node from $G_c$, but the synthetic network $N$ may not achieve this degree for some nodes. Those nodes whose current degree is less than the target degree are said to be ``nodes with available degree". 
At each stage of Step 1,   when adding an edge, we first randomly select nodes within the cluster with available degrees and update their available degree status accordingly. If no suitable nodes with available degree are found, we randomly choose other nodes within the cluster, even if they have no available degree, to add the edge.  We add edges to increase the degree of nodes with available degree, i.e., nodes whose current degree is below their target values, using two different techniques: 

\begin{itemize}
\item  $v1$:  for each node with available degree, we add edges to other nodes with available degree, following Algorithm \ref{algo1}. 
\item $v2$: edges are strategically added to nodes with available degrees, taking into account the number of inter-cluster and intra-cluster edges in the input subgraph $G_c$. 
This approach, detailed in Algorithm \ref{algo2}, restricts the addition of edges to not exceed the expected number of inter-cluster edges. 
\end{itemize}

\begin{algorithm}
\caption{\sc RECCS Stage 4: Version 1}\label{algo1}
\begin{algorithmic}
\Require $length(S) \geq 0$
\State Initialize $available\_node\_degrees \gets \{node: degree \mid node, degree \in S\}$
\State Initialize $max\_heap \gets available\_node\_degrees$

\While{$max\_heap$ is not empty}
    \State $(avail\_degree, available\_node) \gets \texttt{heapq.heappop(max\_heap)}$
    \If{$available\_node \notin available\_node\_degrees$}
        \State \textbf{continue}
    \EndIf
    \State Find $available\_non\_neighbors \gets \{node \mid node \neq neighbor\_of\_available\_node\}$
    \State $avail\_k \gets \min(avail\_degree, \texttt{len}(available\_non\_neighbors))$
    \For{$i \in range(avail\_k)$}
        \State $random\_node \gets \texttt{random}(available\_non\_neighbors)$
        \State Add edge $(available\_node, random\_node)$
        \State Delete $random\_node$ from $available\_non\_neighbors$
        \If{$available\_node\_degrees[random\_node] > 1$}
            \State $available\_node\_degrees[random\_node] \mathrel{-=} 1$
        \Else
            \State Delete $random\_node$ from $available\_node\_degrees$
        \EndIf
    \EndFor
    \State Delete $available\_node$ from $available\_node\_degrees$
\EndWhile
\end{algorithmic}
\end{algorithm}

\begin{algorithm}
\caption{\sc RECCS  Stage 4: Version 2}\label{algo2}
\begin{algorithmic}
\Require Empirical graph $G$, synthetic network $N_{intermediate}$ (SBM-generated network modified by {\sc RECCS} Stages 1-3), clustering $\mathcal{C}$
\Ensure Final synthetic network $N_{output}$
\State Initialize: Read $G$, $N_{intermediate}$, and $\mathcal{C}$
\State Compute inter-cluster and intra-cluster edge probabilities matrix $probs_{\text{emp}}$ based on $G$ and $\mathcal{C}$
\State Initialize $available\_node\_degrees \gets \{node: degree \mid node, degree \in S\}$
\State Compute synthetic edge probabilities matrix $probs_{\text{syn}}$ for $N_{intermediate}$
\State Compute difference matrix $\Delta P = probs_{\text{emp}} - probs_{\text{syn}}$
\State Restrict diagonal and negative values in $\Delta P$ to zero
\State Identify clusters with no $available\_nodes$ and set the corresponding rows and columns in $\Delta P$ to zero
\State Compute $available\_nodes \gets \{available\_node\_degrees.keys()\}$
\State Compute degree sequence $\mathcal{D} \gets \{available\_node\_degrees.values()\}$
\State Compute cluster assignment $\mathcal{C'}$ for $available\_nodes$
\State Generate new graph $N'$ using SBM with $\mathcal{C'}$, $\Delta P$, and $\mathcal{D}$ as input
\State Remove parallel edges and self-loops in the generated network $N'$
\State $N_{output} \gets N_{intermediate} + N'$
\end{algorithmic}
\end{algorithm}

\subsubsection{Comparison of {\sc RECCS}v1 and {\sc RECCS}v2 on development data}

We compare {\sc RECCS}v1 and {\sc RECCS}v2 restricted to Step 1 (i.e., modeling the clustered subnetwork) in Figure \ref{fig_ce_boxplot} with respect to  the metrics discussed in Section \ref{eval_criteria} on the 110 clustered real-world networks in the development dataset. 
Because we only perform Step 1, the evaluation is performed with respect to reproducing features of the clustered subnetwork. 

Both versions of {\sc RECCS} have a better fit than SBM for the minimum edge cuts sequence, showing that the main goal of {\sc RECCS} is achieved well by both variants.  For the other criteria, the differences between the three simulators are too small to be considered meaningful.

\subsection{Experiment 3 results: Evaluation on test data}\label{sec:exp3}

In this experiment, we compare  {\sc RECCS}v1, {\sc RECCS}v2, and SBM synthetic networks on test data.  Unlike Experiment 2, these evaluations are performed with respect to the accuracy of the full network, not just the clustered subnetwork.

We first evaluate how well each method reproduces features of the input clustered real-world networks, according to the evaluation criteria. 
This evaluation is done on the two different testing datasets, Testing 1 and Testing 2.

Since an exact copy of the clustered real-world network is not the objective, we also verify that the synthetic networks are different from the real-world networks by calculating the normalized edit distance, which is the number of edge modifications (addition or deletion) needed to transform one network into the other, then normalized by the number of edges in the real-world network. This evaluation is done on the Testing 1 dataset.

\begin{figure}[hbt!]
\centering
\includegraphics[width=1\textwidth]{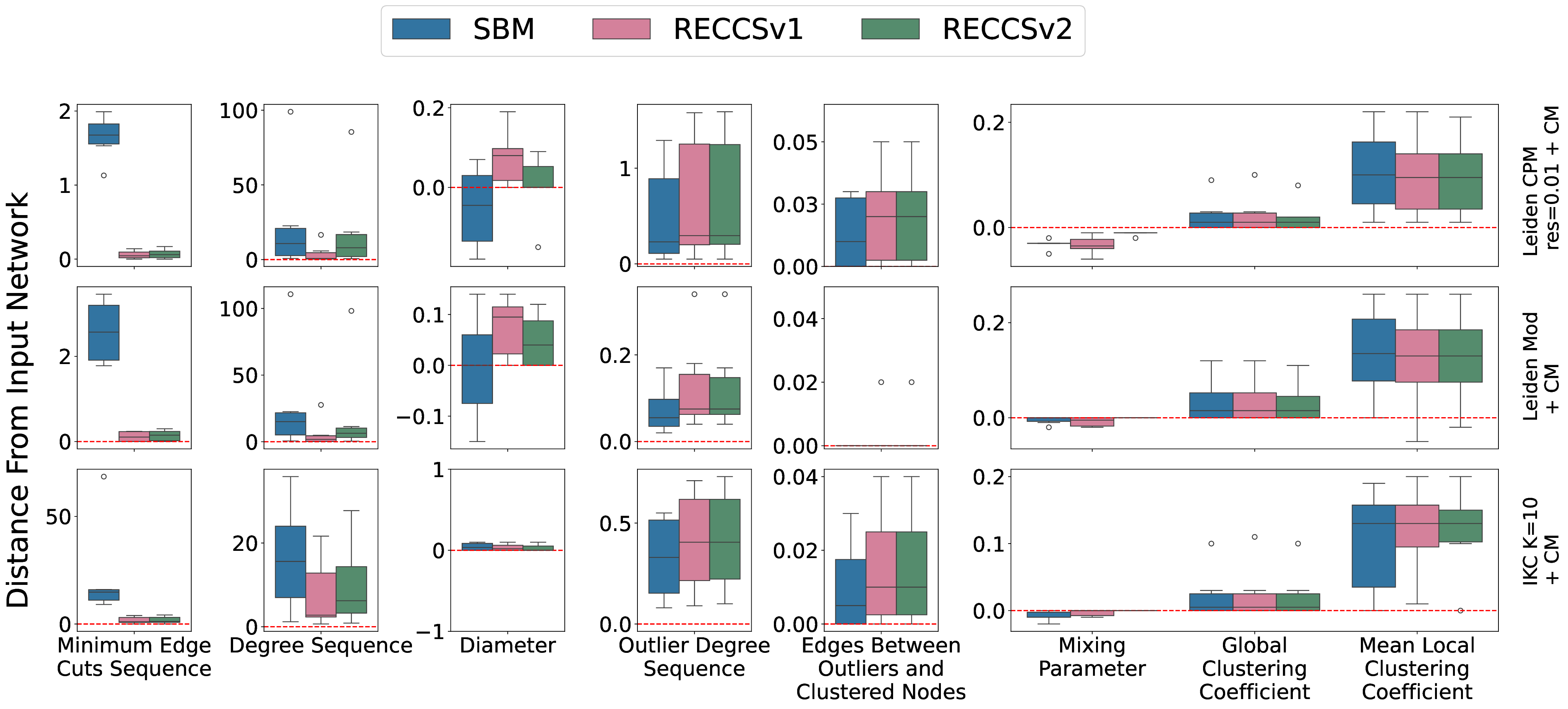}
\caption{\textbf{Experiment 3: Accuracy of SBM and two {\sc RECCS} pipelines on the Testing 1 dataset.} We explore three clusterings:  Leiden-CPM with $r=0.01$ (top row), Leiden-modularity (middle row), and the Iterative k-core (IKC) method (bottom row), each followed by post-processing using CM.}
\label{fig:other-test}
\end{figure}

\paragraph{Accuracy on Testing 1 dataset.}
The clustered networks in the test data, shown in Fig. \ref{fig:other-test}, are based on three different clustering methods, each post-processed to ensure well-connected clusters.  
Across all three clusterings, 
both {\sc RECCS}v1 and {\sc RECCS}v2 demonstrate improvements over SBM in terms of the minimum edge cut size.
This is consistent with observations on the development data, and is expected.

The differences between methods are smaller for the other criteria, but  some are worth noting. Specifically, for degree sequence, {\sc RECCS}v1 has the best fit, followed by {\sc RECCS}v2, and then finally by SBM. 
For diameter, {\sc RECCS}v2 has a better fit than {\sc RECCS}v1, but neither  version is reliably better than SBM.
For outlier degree sequence and number of edges between outliers and clustered nodes, SBM is better than both {\sc RECCS} variants.
For the remaining criteria (i.e., 
mixing parameter, global clustering coefficient, and mean local clustering coefficient) the differences between methods are too small to be meaningful.

Overall, therefore, the three methods are very close for most criteria other than edge cut size, where both {\sc RECCS}v1 and v2 are much better than SBM.

\paragraph{Accuracy on the Testing 2 dataset.}
We next examined results on the Testing 2 dataset.
All clusterings in this set have connected clusters, but in this experiment we compare results when the clustering method is performed with and without the CM post-processing step.

\begin{figure}[hbt!]%
\centering
\includegraphics[width=1\textwidth]{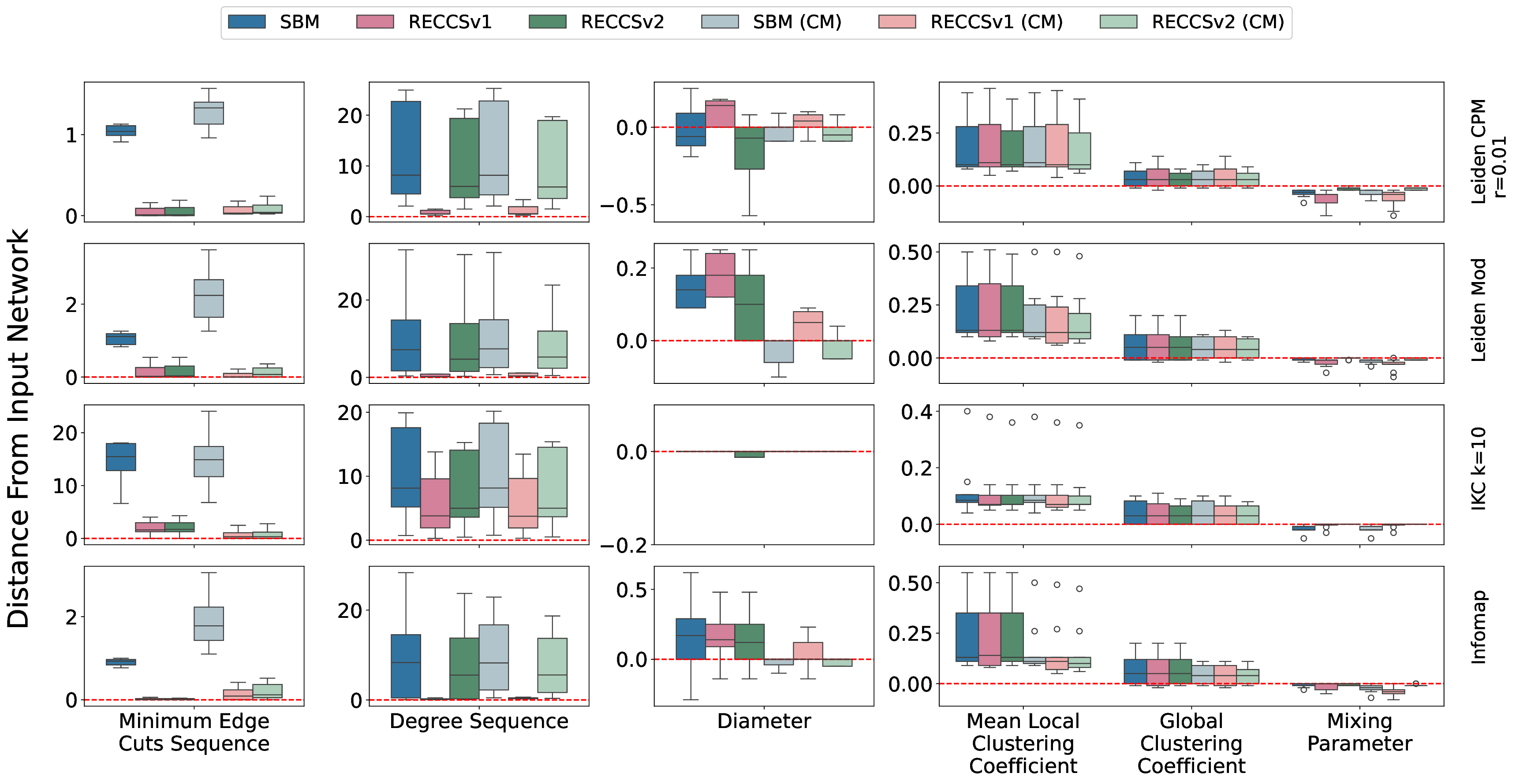}
\caption{\textbf{Accuracy of SBM,   {\sc RECCS}v1, and {\sc RECCS}v2 on the Testing 2 dataset} 
The Testing 2 dataset has nine networks clustered with four methods: Leiden-CPM with $r=0.01$ (top row), Leiden-modularity (second row), the Iterative k-core (IKC) method with $k=10$ (third row), and Infomap (bottom row). The ``CM "in the legend indicates that the input clustering includes a  post-processing step to ensure that the clusters are well-connected.  The y-axis shows different distance metrics for various network properties.
All clustering methods other than IKC returned at least one non-singleton cluster on each real-world network;  however,  IKC k=10 returned 0 clusters for com-Amazon, so that network is excluded from results for IKC. } 
\label{fig:additional-test}
\end{figure}
\clearpage

The results of this experiment, shown in Figure \ref{fig:additional-test}, permit several observations regarding the performance of SBM and the two {\sc RECCS} pipelines across different input clustering methods. 

Across all three basic input clusterings and whether or not post-processed by CM, we again see that both {\sc RECCS}v1 and {\sc RECCS}v2 have a better fit to the minimum edge cut sequence than SBM.
We also see that {\sc RECCS}v1 has a better fit to degree sequence than the other methods, and that {\sc RECCS}v2 is slightly better than SBM.

Results on diameter are mixed, with no method outperforming any other, although possibly {\sc RECCS}v2 can be considered to be the best performing of the three methods. All methods have very similar scores for the local clustering coefficients, global clustering coefficients, and mixing parameters, but with a small advantage to {\sc RECCS}v2 over the other methods.   
When the input clustering parameters come from clusterings post-processed with CM, accuracy measurements generally improve for all methods and criteria, with very dramatic improvements for  diameter and mean local clustering coefficient.
This trend suggest that   networks that have well-connected clusters are easier to model using {\sc RECCS}.

\begin{figure}[hbt!]%
\centering
\includegraphics[width=0.9\textwidth]{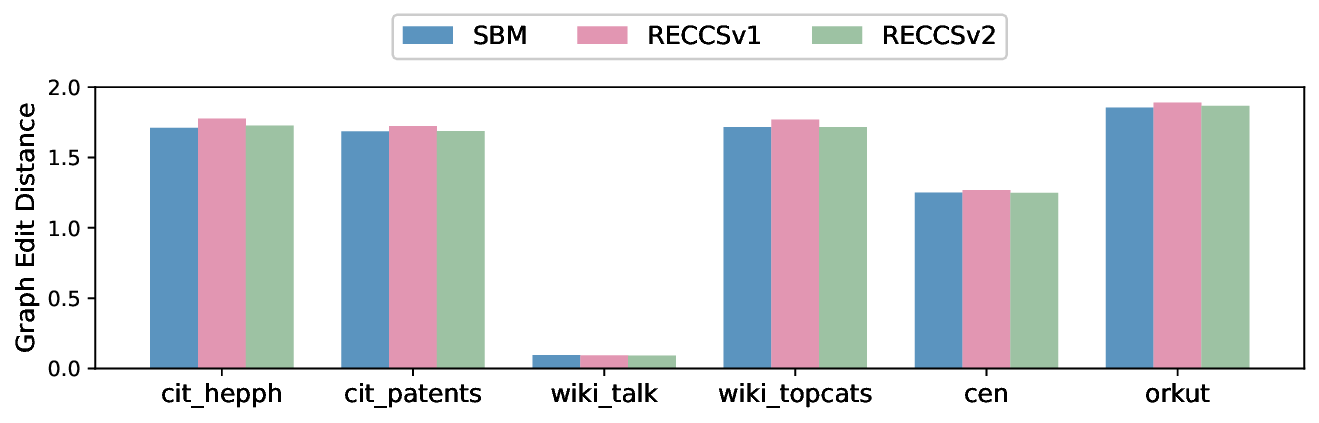}
\caption{\textbf{Comparing SBM, {\sc RECCS}v1,  and {\sc RECCS}v2 with respect to the normalized edit distance between synthetic and real world networks}
The  normalized edit distance between 
the edge sets of the true network $G$ and the synthetic network $N$, i.e., $\frac{|E(G)  \triangle E(N)|} {|E(G)|}$, where $\triangle$ denotes the symmetric difference, and so the maximum possible value is $2.0$. Each real-world network is clustered using
Leiden-CPM, with   $r=0.01$. Here, {\sc RECCS}v2  produces synthetic networks that are closer to the real-world network than {\sc RECCS}v1, and about as close as SBM networks. 
This experiment is on the test data.
}
\label{fig:graph_edit_distance}
\end{figure}

\paragraph{Edit distance to the real-world network}
We evaluated the normalized edit distance between the real-world network and the synthetic networks generated by the two {\sc RECCS} pipelines and the unmodified SBM. This metric calculates the number of edge modifications (additions or deletions) required to transform the real-world network into the synthetic network, normalized by the total number of edges in the real-world network.

For the six test networks, SBM and {\sc RECCS}v2 showed similar performance, while {\sc RECCS}v2 consistently produced synthetic networks with a smaller normalized edit distance compared to {\sc RECCS}v1 (Fig. \ref{fig:graph_edit_distance}). Notably, with a maximum possible normalized edit distance of $2.0$, the results indicate that for all but one network, the generated synthetic networks exhibit sufficient randomness to exclude being labeled as replicates of the input.

In summary, Experiment 3 shows that both RECCS pipelines improve upon SBM in terms of cluster edge connectivity.  Differences between {\sc RECCS}v1, {\sc RECCS}v2, and SBM are generally small for the other criteria, but some differences between the two {\sc RECCS} variants are clear. 
Specifically, {\sc RECCS}v1  has better accuracy with respect to reproducing the degree sequence, while {\sc RECCS}v2 performs better for modeling diameter and the mixing parameter. 
Finally, all three methods produce networks that differ from the six Testing 1 real-world networks, as evidence from the graph edit distance, but the differences range from very small on the wiki\_talk network to relatively large on the other networks we studied. 

\subsection{Experiment 4 results: Comparing clustering methods on RECCS networks}
\label{sec:exp4}

This experiment evaluates clustering methods in terms of ability to find the ground truth clusters in {\sc RECCS}v1 networks.

We use the Testing 2 dataset for this experiment, which includes nine real-world networks clustered using four different clustering methods, but without CM post-processing.
These input parameters are given to {\sc RECCS}v1 to create the synthetic ground truth clusters.

Each synthetic {\sc RECCS} network is clustered using eight different clustering methods, defined by the same four basic techniques (Leiden-CPM(0.01), Leiden-mod, IKC, and Infomap), but in this case each with or without CM post-processing. 

\begin{figure}[hbt!]%
\centering
\includegraphics[width=1\textwidth]{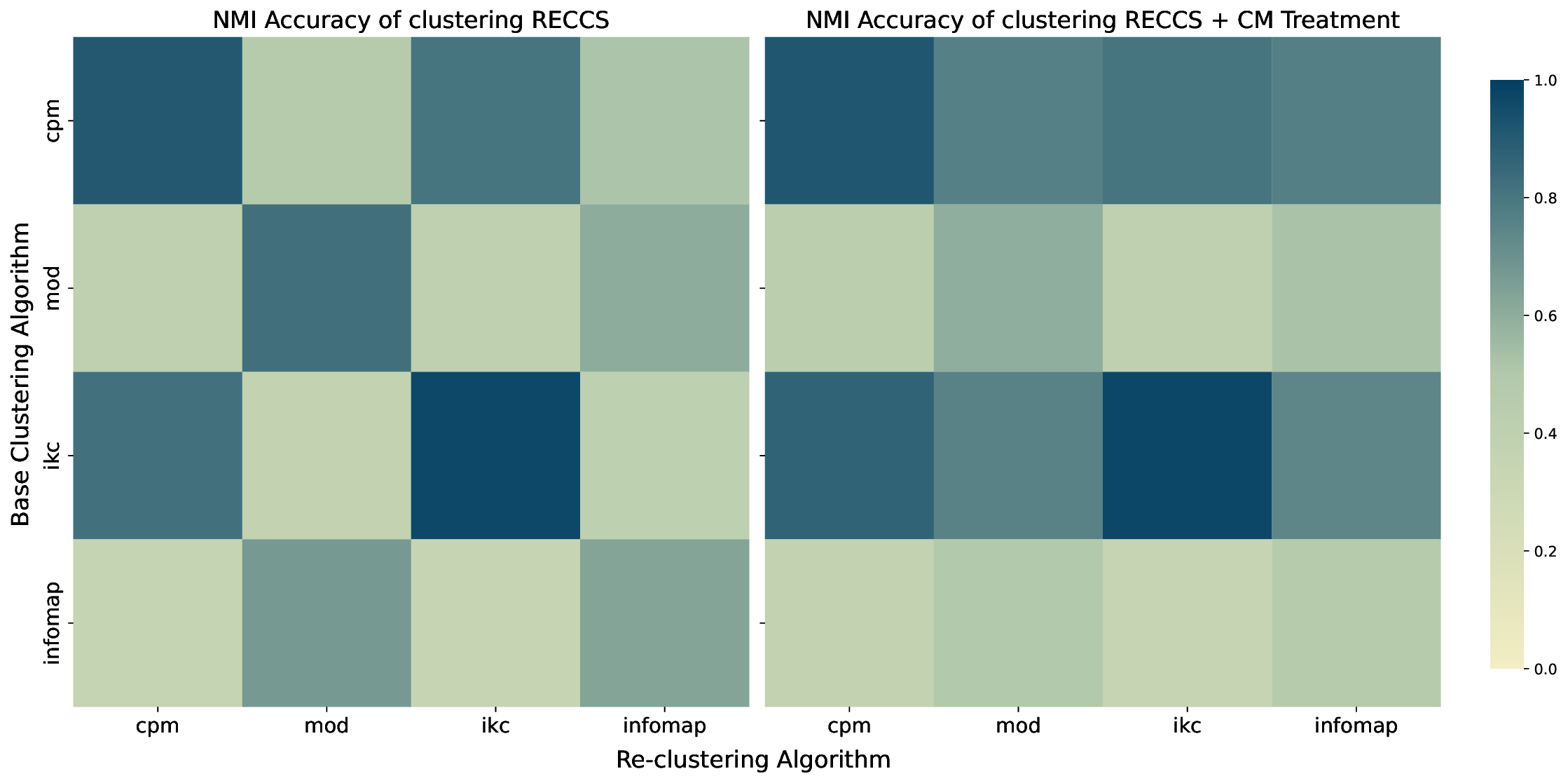}
\caption{\textbf{Accuracy (NMI) of clustering methods on RECCS Networks.} Average Normalized Mutual Information (NMI) scores
for re-clustered synthetic networks generated by RECCS are shown in this heatmap. Rows correspond to the base clustering algorithm used to derive the input parameters for RECCS. The columns represent the re-clustering algorithms applied. The left heatmap shows results for clustering without CM treatment, while the right heatmap corresponds to CM-treated re-clusterings. Higher NMI scores, indicated by darker shades, signify stronger alignment between re-clustered networks and the ground truth clustering. IKC k=10 returned 0 clusters for "com-Amazon", which was consequently excluded from computing the average for IKC. These results are performed using Testing 2 datasets.}
\label{fig:recluster-nmi}
\end{figure}

\begin{figure}[hbt!]%
\centering
\includegraphics[width=1\textwidth]{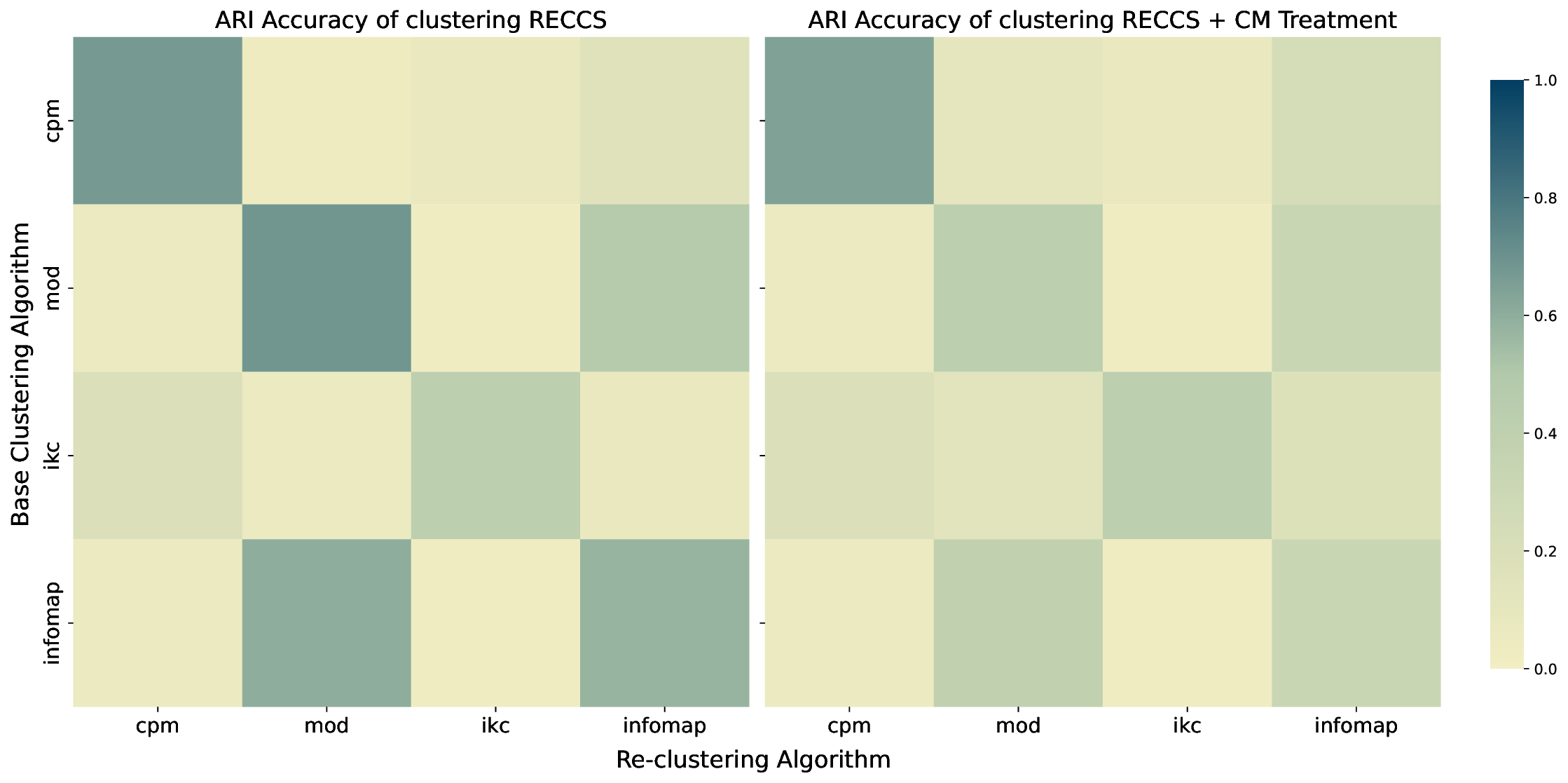}
\caption{\textbf{Accuracy (ARI) of clustering methods on RECCS Networks with or without CM} Adjusted Rand Index (ARI) scores are shown for re-clustered synthetic networks generated by {\sc RECCS} are shown in this heatmap. Rows correspond to the base clustering algorithm used to derive the input parameters for {\sc RECCS}. Columns represent the  algorithms used to cluster the synthetic {\sc RECCS} network. The left panel shows results for clustering without CM treatment, while the right panel corresponds to CM-treated re-clusterings. Higher ARI scores, indicated by darker shades, signify stronger alignment between re-clustered networks and the ground truth clustering.}
\label{fig:recluster-ari}
\end{figure}

Accuracy with respect to recovering the ground truth communities is assessed using NMI (Figure \ref{fig:recluster-nmi}) and ARI (Figure \ref{fig:recluster-ari}).
In each figure, the left subfigure shows results when the clustering method used to cluster the {\sc RECCS} network is performed without the CM post-processing treatment, and the right subfigure is
for the case where the clustering is post-processed by CM. 
The rows in each figure correspond to the base clustering method (i.e., Infomap, IKC, Leiden-mod, and Leiden-CPM) used to provide input parameters to {\sc RECCS}.
The columns in each figure also correspond to the base clustering methods used to re-cluster the {\sc RECCS}v1 network.
Thus, each square in each subfigure defines the pair of clustering methods, where the row indicates which method is used for parameter generation and the column indicates which method is used to cluster the synthetic network. Hence, squares down the diagonal in the left subfigure indicate that the same method is used in each case; squares down the diagonal in the right subfigure are  for the case where the method used to re-cluster {\sc RECCS} network is followed by CM treatment, although the method used to compute input to {\sc RECCS} is not.

The re-clustering experiment provides the following insights, as illustrated in Figures \ref{fig:recluster-nmi} and \ref{fig:recluster-ari}:

\begin{itemlist}
\item \emph{Fidelity with Input Clustering:} Both NMI (Figure~\ref{fig:recluster-nmi}) and ARI (Figure \ref{fig:recluster-ari})   scores are highest when the re-clustering algorithm matches the original clustering algorithm used to compute the input parameters for {\sc RECCS}. For instance, Leiden-CPM as the base clustering produces the best scores when re-clustered with Leiden-CPM.

\item \emph{Similar outcomes for pairs:}  In these experiments, Leiden-mod and Infomap each do well on the same model conditions, and similarly  Leiden-CPM(0.01) and IKC each do well on the same model conditions.  

\item \emph{Impact of CM Treatment:} Using CM post-processing often improves  NMI clustering accuracy but not ARI accuracy.  
\end{itemlist}

\begin{table}[ph]
\tbl{Replication experiment of {\sc RECCS} on the Curated Exosome Network (CEN) using Leiden-CPM(0.01) clustering. \label{tbl:replication-res}}
{\begin{tabular}{@{}ccccc@{}} \toprule
\makecell{Input \\ Post-processing} & Network Property & SBM & \makecell{RECCSv1 \\ (Mean ± Std)} & \makecell{RECCSv2 \\ (Mean ± Std)} \\ \colrule
\multirow{7}{*}{\makecell{With CM \\(without \\ filtering \\small \\ clusters)}}
& Degree Sequence & 16.2516 & 3.70 ± 0.01 & 11.63 ± 0.01 \\
& Diameter & -0.2308 & -0.10 ± 0.04 & -0.23 ± 0.00 \\
& Global Clustering Coefficient & 0.0005 & 0.00 ± 0.00 & 0.00 ± 0.00 \\
& Mean Local Clustering Coefficient & 0.0147 & -0.00 ± 0.00 & -0.00 ± 0.00 \\
& Minimum Edge Cuts Sequence & 1.2499 & 0.00 ± 0.00 & 0.00 ± 0.00 \\
& Mixing Parameter & -0.0339 & -0.02 ± 0.00 & -0.01 ± 0.00 \\
& Outlier Degree Sequence & 0.1025 & 0.19 ± 0.00 & 0.19 ± 0.00 \\ \colrule
\multirow{7}{*}{\makecell{No  \\ post-processing \\ \\}} 
& Degree Sequence & 23.0437 & 1.36 ± 0.00 & 19.79 ± 0.01 \\
& Diameter & -0.3846 & -0.31 ± 0.00 & -0.36 ± 0.04 \\
& Global Clustering Coefficient & 0.0002 & -0.00 ± 0.00 & 0.00 ± 0.00 \\
& Mean Local Clustering Coefficient & 0.0482 & 0.04 ± 0.00 & 0.04 ± 0.00 \\
& Minimum Edge Cuts Sequence & 1.0839 & 0.00 ± 0.00 & 0.00 ± 0.00 \\
& Mixing Parameter & -0.0476 & -0.02 ± 0.00 & -0.00 ± 0.00 \\
& Outlier Degree Sequence & 0.0680 & 0.13 ± 0.00 & 0.13 ± 0.00 \\ \botrule
\end{tabular}}
\begin{tabnote}
This table presents the comparison of SBM and RECCS network metrics across various network properties, with and without CM post-processing for input clustering. RECCSv1 and RECCSv2 values represent the mean and standard deviation (Mean ± Std) of three replicates generated by running RECCS on the same SBM base network.
\end{tabnote}
\end{table}

\subsection{Experiment 5 results: Replication}
\label{sec:exp5}

The replication experiment evaluates the consistency and robustness of the {\sc RECCS} framework in generating synthetic networks.
For this experiment, the real-world Curated Exosome Network (CEN) with its 13,989,436 nodes and 92,051,051 edges was used as a single benchmark.
We cluster the CEN using just one clustering method (Leiden-CPM(0.01)), with and without CM post-processing. 
For each SBM base network, {\sc RECCS} was applied three times, and variability in the  accuracy values was calculated. 

Table \ref{tbl:replication-res} presents the results for key  properties across synthetic networks generated by SBM, {\sc RECCS}v1, and {\sc RECCS}v2, with and without CM post-processing for input clustering.
The  standard deviations for {\sc RECCS}v1 and {\sc RECCS}v2 are very small across all network properties, demonstrating that the synthetic network quality measures are reliably similar between replicates on this input condition.  

\subsection{Experiment 6 results: Runtime for RECCS}
\label{sec:exp6}

This section analyzes the runtime  of the {\sc RECCS} framework on the  largest Testing 2 network, com-LiveJournal, which has
3,997,962 nodes and 34,618,189 edges.
Four different clustering methods were used, each with and without  CM post-processing.
Runtimes in minutes are shown   in Table~\ref{tbl:runtime}. 

\begin{table}[ph]
\tbl{Runtime (mins) for SBM and RECCS network generation with different input clusterings, with or without CM post-processing, on com-LiveJournal ($\sim$4M nodes and $\sim$35M edges). \label{tbl:runtime}}
{\begin{tabular}{@{}cccccc@{}} \toprule
Clustering & \makecell{SBM} & \makecell{RECCSv1} & \makecell{RECCSv2} & \makecell{Node \\ Coverage \\ (\%)} & \makecell{Number \\ of \\ Clusters} \\ \colrule
Leiden-CPM(0.01) & 21 & 319 / 3 & 255 / 3 & 92.07 & 148876 \\
Leiden-Mod & 20 & 81 / 1 & 119 / 1 & 100.00 & 2006 \\
IKC k=10 & 44 & 9 / 38 & 19 / 38 & 10.60 & 3163 \\
Infomap & 20 & 48 / 1 & 85 / 1 & 100.00 & 2 \\ \botrule
\hline
Leiden-CPM(0.01) +CM & 20 & 191 / 4 & 139 / 4 & 70.65 & 153353 \\
Leiden-Mod +CM & 22 & 36 / 8 & 55 / 8 & 44.34 & 12835 \\
IKC k=10 +CM & 43 & 10 / 37 & 20 / 37 & 10.52 & 4560 \\
Infomap +CM & 20 & 37 / 7 & 48 / 7 & 38.78 & 16745 \\ \botrule
\end{tabular}}
\begin{tabnote}
The column ``SBM" represents the time taken to generate the entire SBM network using graph-tool under the degree-corrected model. The ``RECCSv1" and ``RECCSv2" columns present the runtime for Step 1 and Step 2 of {\sc RECCS} as ``Step 1 / Step 2". Step 1 includes the SBM computation on the clustered subnetwork, while Step 2 accounts for adding the outliers. ``Node coverage" refers to the percentage of nodes in non-singleton clusters, and ``Number of Clusters" represents the total count of non-singleton clusters.
\end{tabnote}
\end{table}

The runtimes for both {\sc RECCS} versions vary significantly, and depend on the input clustering method, and whether CM post-processing is used. 
Of note, the comparison between the {\sc RECCS} pipelines shows that typically {\sc RECCS}v1 is slower than  {\sc RECCS}v2, but this is not always the case. For example, on one of the eight input conditions, {\sc RECCS}v2 is slower.

When the clusterings are not post-processed with CM,   runtimes range from 47 minutes to 322 minutes  for {\sc RECCS}v1 and from 57 to 259 minutes  for {\sc RECCS}v2.
However, when clusterings are post-processed with CM, runtimes are faster, ranging from 44 minutes to 195 minutes for {\sc RECCS}v1 and from 55 to 143 minutes for {\sc RECCS}v2. 
Thus,  CM post-processing the input clustering reduces the time for both {\sc RECCS} versions on all clusterings.

The results show that {\sc RECCS} scales effectively to large networks in the range of com-LiveJournal.

\section{Conclusions}

Synthetic networks with planted ground truths are a useful approach for evaluating community detection methods. Of these, SBM network generation implemented in graph-tool  is a valuable tool for constructing synthetic networks that mimic real world networks. In examining the properties of ground-truth clusters generated by SBMs, we observed a significant number of cases where these clusters were disconnected.  In this study, the input to graph-tool for SBM construction was a real-world network clustered with different community detection algorithms, and the trend of disconnected clusters was seen even under conditions where the input clustering was guaranteed to be connected. 

We developed {\sc RECCS}, a meta-method, to remediate disconnected clusters. {\sc RECCS}  is embedded in a pipeline that takes a real world network and a user-specified clustering, generates an SBM and modifies it to return a synthetic network complete with outlier nodes, node-to-node and cluster-to-cluster correspondence with the input, and comparable edge-connectivity. {\sc RECCS}  introduces randomness into intra-cluster and inter-cluster edges, thus achieving stochastic variance from the input.  

We demonstrate that {\sc RECCS}  is able to restore the minimum edge cuts of DC-SBM generated clusters to make them comparable to the input clusters.  
In our testing of {\sc RECCS}, we demonstrate scalability to networks in the order of 14 million nodes. We also report on the replicability of results from {\sc RECCS} networks, showing that network quality measures are reliably consistent between replicates.

We stress that {\sc RECCS} is empirically motivated.  In addition, we  argue that {\sc RECCS} provides value to practitioners in different domains who are presented with the challenge of identifying a suitable clustering technique for their investigations.  {\sc RECCS} networks are essentially stochastic variants of a given real world network for which ground truths and comparable networks do not exist. We argue that all networks are not similar, therefore using a set of benchmark networks of diverse origin to evaluate a clustering method is useful.

 \section*{Acknowledgments}
The authors acknowledge the Insper-Illinois partnership for funding that partially supported this work.



 \section*{ORCID}
 \noindent LA: \url{https://orcid.org/0009-0007-5708-2405}\\
 \noindent TVL: \url{https://orcid.org/0000-0002-4480-5535} \\
 \noindent MP: \url{https://orcid.org/0000-0002-8676-7565}\\
\noindent TW: \url{https://orcid.org/0000-0001-7717-3514} \\
 \noindent GC: \url{https://orcid.org/0000-0002-2127-1892}





 \appendix


\clearpage
\bibliographystyle{ws-acs}
\bibliography{main}

\end{document}